# An explainable Recursive Feature Elimination to detect Advanced Persistent Threats using Random Forest classifier

Noor Hazlina Abdul Mutalib
Department of Artificial Intelligence,
Faculty of Computer Science and Information Technology
University Malaya, Malaysia
s2121830@siswa.um.edu.my

Aznul Qalid Md Sabri
Department of Artificial Intelligence,
Faculty of Computer Science and Information Technology
University Malaya, Malaysia
aznulqalid@um.edu.my

Ainuddin Wahid Abdul Wahab
Department of Computer System & Technology, Faculty of Computer Science and Information Technology
University Malaya, Malaysia
ainuddin@um.edu.my

Erma Rahayu Mohd Faizal Abdullah
Department of Artificial Intelligence,
Faculty of Computer Science and Information Technology
University Malaya, Malaysia
erma@um.edu.my

Nouar AlDahoul
Department of Computer Science,
New York University Abu Dhabi, Abu Dhabi, UAE
naa9497@nyu.edu

*Abstract*— Intrusion Detection Systems (IDS) play a vital role in modern cybersecurity frameworks by providing a primary defense mechanism against sophisticated threat actors. In this paper, we propose an explainable intrusion detection framework that integrates Recursive Feature Elimination (RFE) with Random Forest (RF) to enhance detection of Advanced Persistent Threats (APTs). By using CICIDS2017 dataset, the approach begins with comprehensive data preprocessing and narrows down the most significant features via RFE. A Random Forest (RF) model was trained on the refined feature set, with SHapley Additive exPlanations (SHAP) used to interpret the contribution of each selected feature. Our experiment demonstrates that the explainable RF-RFE achieved a detection accuracy of 99.9%, reducing false positive and computational cost in comparison to traditional classifiers. The findings underscore the effectiveness of integrating explainable AI and feature selection to develop a robust, transparent, and deployable IDS solution.

INTRODUCTION

Cyberattacks classified as Advanced Persistent Threats (APTs) are distinguished by their precision, advanced techniques, and prolonged purposes. APTs are meticulously planned to exploit system vulnerabilities and can remain undetected for extended periods [1]. Their objective is often to continuously monitor, exfiltrate, or manipulate data across multiple stages of attack. Attackers employ evolving techniques such as polymorphic malware, encrypted command-and-control channels, and the masking of malicious behavior within legitimate network traffic. These tactics make detection challenging and diminish the efficiency of conventional and signature-based IDS, which rely heavily on predefined attack patterns [2][3]. These attacks involve multiple phases, like reconnaissance, lateral movement, privilege escalation, and data exfiltration. This study delves into the critical role APTs play in today's threat landscape, uncovers current research limitations, and considers the broader implications for developing more resilient cybersecurity solutions.

Developing an effective mechanism to detect APTs presents several significant challenges. One of the foremost issues is handling the high dimension of network traffic data, includes hundreds of flow-based features. This issue can lead to overfitting and excessive computational demands, making it difficult for conventional machine learning models to perform efficiently [4]. Another major hurdle is the inherent class imbalance in real-world datasets, where benign traffic significantly outnumbers malicious activity. As a result, models tend to favour the majority class, thereby reducing their sensitivity actual threats. In addition, many cutting-edge ML classifiers operate as black-box systems, offering minimal insight into their internal reasoning [1],[21]. This lack of interpretability undermines trust and complicates security incident responses.

## I. RELATED WORK

Several studies have explored ML and DL approaches to improve threat detection in response to the evolving threat landscape posed by APTs. For example, RF classifiers have demonstrated robust and strong performance in handling high-dimensional and class-imbalanced datasets [5].

Feature selection methods such as RFE are employed to boost model accuracy and interpretability by identifying the most informative features and reducing redundancy [6]. Prior research has also underscored the importance of model explainability in cybersecurity, aiming to deliver actionable insights for cybersecurity analysts. However, few studies have specifically combined explainable feature selection techniques with ensemble classifiers for APT detection. Recent works [4], [21-25] have progressively focused on merging explainability into IDS to enhance their trustworthiness and interpretability.

In [7], the authors introduced a new stable feature selection algorithm called IV-RFE, which includes relative variance and weighting in RFE for improved intrusion detection. This method enhances feature relevance and stability. Their model was evaluated on UNSW-NB15 dataset across three attack types: reconnaissance, analysis, and denial of service (DoS). Their results showed that IV-RFE improved accuracy and stability but also achieved up to 98.6% accuracy with a stability index of 0.98 in detecting DoS attacks.

In [8], the author proposed the Multi-strategy RIME Optimization Algorithm (MRIME) in NIDS. MRIME enhances the original RIME algorithm by integrating three key strategies: a chaotic local search to improve exploration, an interaction system to intensify communication between search agents, and a puncture mechanism to prevent premature convergence. The algorithm was evaluated on three public datasets: CICIoV2024, CIC-IDS-2017, and UNSW-NB15. The outcomes demonstrated MRIME's capability to reduce feature dimensions while improving detection performance, making it suitable for high-dimensional and large-scale datasets. However, this study focused on the RF classifier, limiting its generalizability against baseline models.

In [9], the authors proposed and tested a Stepwise Recursive Feature Elimination (SRFE) to enhance NIDS, which incrementally removes fewer notable features based on Information Gain (IG). The authors tested SRFE on the NSL-KDD dataset using RF, J48, SVM, Naive Bayes, and the classifiers performed well. RF and J48 achieved the highest accuracy, up to 99.80% for binary classification and 99.69% for multiclass detection. SRFE effectively reduces high-dimensional data, making machine learning models faster and more accurate for intrusion detection.

In [10], the authors presented a feature selection-based approach to enhance IDS performance by utilizing RFE to reduce irrelevant data and improve classification accuracy. The study made use of the UNSW-NB15 dataset using multiple ML classifiers. They applied RFE to manage the high dimensionality of the data, which helped improve classification accuracy and reduced computational load.

In [11], the authors addressed cybersecurity threats faced by smart healthcare systems on the Internet of Medical Things (IoMT) during COVID-19 pandemic. They introduced an anomaly intrusion detection framework using RFE, machine learning models, and ridge regression within a DL framework. The aim was to enhance the detection of abnormal network behavior and protect sensitive medical data. This experiment was conducted on the WUSTL-EHMS dataset. Their RFE-based decision tree model achieved 99% accuracy during training, 97.85% accuracy on testing, and a low false alarm rate (FAR) of 0.03, showing potential for securing IoMT environments.

In [12], the authors evaluated the effect of RFE, Mutual Information (MI), and Lasso Feature Selection (LFS) on boosting Cyber Anomaly Detection Systems (CADS) performance. They proposed a hybrid ensemble classification approach including Random Forest, XGBoost, Extra-Trees, and a logistic regression meta-classifier. Their goal was to identify which method best streamlines relevant features to improve detection accuracy.

[13] analyzed the impact of batch reinforcement learning on correlation-based information gain (IG) and RFE methods on detecting various cyberattacks in deep reinforcement learning (DRL) network environments. The study demonstrated significant improvements in detection accuracy.

[14] presented a generalizable computational framework that evaluated nine data-level augmentation techniques. They evaluated SMOTE, SMOTE-ENN, Borderline-SMOTE, SVM-SMOTE, ADASYN, RUS, ROS, and CT-GAN along with nine (9) ensemble models such as LightGBM, AdaBoost, XGBoost, Gradient Boosting, RF, Voting-Soft, Voting-Hard, Stacking-I, and Stacking-II. There were tests on 23 benchmark datasets with varying imbalance ratios. They also shared their code and data openly to support transparency and reuse. Through 30 independent runs per combination reporting mean, best, and standard deviation of F1, AUC, and accuracy, their study demonstrated that traditional oversampling approaches, SMOTE coupled with LightGBM or SMOTE with the Stacking-II architecture, achieved the highest F1-macro and AUC scores overall.

[15] proposed a model feature selection method that combines feature screening with Random Forest-based Recursive Feature Elimination (RF-RFE) to efficiently handle high-dimensional data. Unlike traditional methods that rely on model assumptions, this approach works flexibly across different data structures and statistical models. This approach has some solid theoretical properties, specifically when it comes to select a relevant feature set and maintaining L2 consistency. This experiment achieved superior performance in simulations on Tecator and Daily Demand Orders datasets. That kind of versatility is what makes it practical beyond just academic use.

A recent study by [16] proposed a hybrid IDS technique based on Decision Tree-Recursive Feature Elimination (DT-RFE) combined with ensemble learning. The approach employed decision tree-based RFE algorithm to reduce feature dimensionality and eliminate irrelevant data, enhance resource efficiency, and lower time complexity. Integrating DT-RFE with a stacking ensemble model significantly improved detection performance. This study demonstrated that the model achieved over 99% accuracy in most attack categories using KDD CUP 99 and NSL-KDD datasets. This confirmed the model effectiveness in providing precise and reliable intrusion detection.

In [17], the authors proposed hybrid learning approaches that integrate feature selection methods, which are univariate information gain ranking and multivariate correlation-based filtering with imbalance learning methods such as Random Under-Sampling, SMOTE over-sampling, MinCost decision rules, and instance weighting to improve RF classification performance measured by F-measure and G-mean on high-dimensional and class-imbalanced datasets.

## II. METHODS

The proposed methodology is structured into several key phases: data preprocessing, feature selection through RF-RFE, classification using ML and DL models, and explainability with SHAP analysis and evaluation on test datasets. This pipeline is specifically designed to address the complexities of detecting APTs within high-dimensional network traffic while ensuring model transparency and robustness. In this section, we break down each step involved in transforming raw network traffic data into an interpretable and effective model for APT detection.

Fig. 1 illustrates the proposed architecture of APT detection. The framework uses the CICIDS2017 dataset. The process begins with raw network traffic flows that are subjected to data preprocessing, including imputation, normalization, and label encoding. Subsequently, RFE is used to choose and retain the most informative features. The trained model categorizes network traffic as benign or an attack and identifies the most influential features contributing to the predictions. Finally, explainable AI (XAI) is employed to ensure transparency, and build trust in the model's decisions.

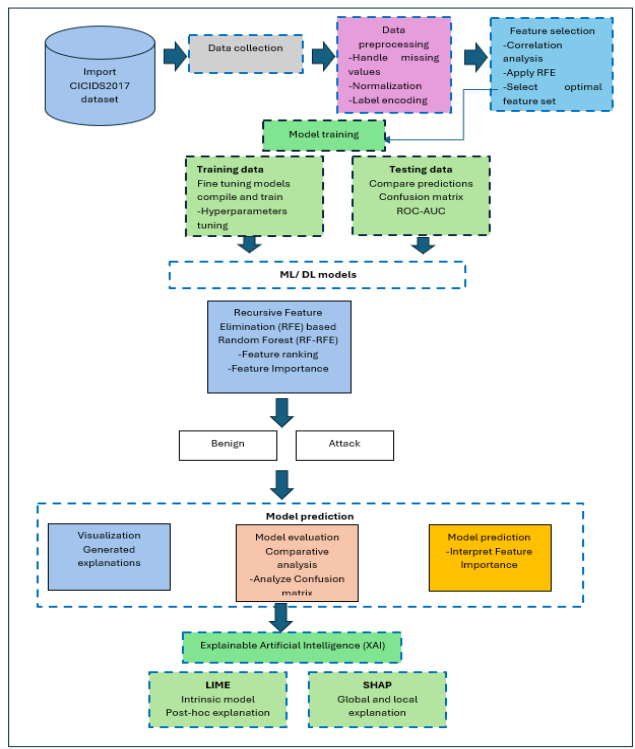

Fig. 1. System architecture of our proposed framework

### A. Dataset

In this study, we evaluate our framework using the CICIDS2017 dataset, a publicly available benchmark containing 2,830,743 labeled network flow data. It includes one benign class and traffic from seven attack types: brute force FTP/SSH, DoS/DDoS, Heartbleed, web-based exploits, infiltration, botnet activity, and port scans [12],[25]. This dataset contains 80 features extracted from flow metadata, including packet sizes, session, protocol flags, host with binary labels [31]. We selected CICIDS2017 for its comprehensive coverage of modern attack vectors and fine-grained labeling, which supports rigorous model validation in real-world IDS environments.

To address the class imbalance, where benign samples significantly outnumber attack samples, we applied stratified sampling and used evaluation metrics such as the F1-score to balance precision and recall. Although we explored SMOTE and class weighting, we found that feature selection via RFE inherently improved the model's sensitivity to minority classes without requiring extensive resampling.

### B. Feature selection with RFE

Feature selection is an important part of developing effective IDS systems, especially when dealing with high-dimensional datasets in network security [18]. Such datasets have many irrelevant, noisy, or redundant features. Irrelevant features that do not contribute to anything might reduce accuracy. Noisy features contain inconsistencies or errors that could damage the reliability of any model built over such data. When building a detection model, the most important component is narrowing things down to the most meaningful features. Removing redundant computations enhances model efficiency, reduces execution time, and simplifies interpretability [19].

In this research, RFE was employed with a Random Forest (RF) classifier to identify the most informative features while reducing model complexity [20]. Initially, all 80 features were ranked based on impurity scores from the RF estimator. The least important features were iteratively eliminated until the model achieved optimal performance with 20 top-ranked features, maintaining an F1-score above 0.90. The final feature set was further validated using SHAP to ensure both statistical relevance and interpretability. Selected features such as Flow Duration, Total Forward Packets, Bwd Packet Length Std, and Avg Bwd Segment Size proved effective in capturing subtle traffic patterns indicative of APTs.

### C. Random Forest (RF) classifier

Random Forest (RF) is an ensemble learning algorithm that constructs multiple decision trees to enhance prediction accuracy and reduce overfitting [21]. Each tree is trained on a random subset of the data and features. This improves classification accuracy and reduces bias. The feature importance scores created by RF model can be used to determine the most influential features in a dataset.

In supervised learning, the goal of RF is to predict an outcome, denoted as $Y_i$ for each sample i (where i=1,…,n) using predictor variables and $X_{ij}$ for (where j=1,…, P) variables. At each split, every variable has an equal chance of being selected as a candidate. RF computes feature-importance scores to identify and remove irrelevant features, which helps improve generalization on large feature sets. RF is chosen as a classifier in many applications due to its robustness, high accuracy, effectiveness on large-scale, and imbalanced datasets.

### D. Explainable Artificial Intelligence (XAI) with SHAP

Explainability is a crucial aspect for modern IDS, specifically those relying on machine learning models [22]. While machine learning models often yield accurate predictions, their decision-making processes are frequently opaque, making it difficult for cybersecurity analysts to understand why a particular activity has been flagged as suspicious [23]. As a result, this lack of transparency can cause distrust in the model's predictions and hinder cybersecurity analysts' ability to take appropriate action based on identified threats [23].

SHapley Additive exPlanations (SHAP) is a technique that explains model outputs by assigning Shapley value to each feature representing its average contribution to the prediction across all possible feature combinations [24]. SHAP values can be used to comprehend the overall importance of each feature and its contribution to individual predictions. This information identifies the most influential features and clarifies why a model made a particular prediction for a specific instance. SHAP generates feature importance rankings, summary plots, and decision plots to visualize how specific attributes influence the model's output [32]. This level of explainability is vital for cybersecurity analysts, enabling them to trace alerts back to specific traffic behaviors and enhancing trust in the detection system.

### E. Experimental setup

We split the fully pre-processed CIC-IDS2017 dataset that was divided into 80:20 percent using stratified sampling to ensure that the original benign-to-anomaly ratio is correct [26]. The experimental setup was conducted on a workstation installed with an Intel Core i7-8650U CPU, 16 GB RAM, and Intel UHD Graphics. All models were developed and evaluated using Python 3.8.10. The implementation utilized the following libraries:

*Table 1. Hardware and software utilized for experiment setup.*

| No. | Components | Parameters |
|---|---|---|
| 1 | Processor | Intel Core i7-8650U CPU |
|   | Installed RAM | 16 GB RAM |
| 2 | Operating system | Window 11 Pro |
| 3 | Programming | Python 3.8.10 |
| 4 | Deep learning package | TensorFlow 2.17.0, Scikit-Learn 1.5.1, Pandas 2.2.2, NumPy 1.26.4 |
| 5 | Environment | Google Colab |

## III. RESULTS ANALYSIS

This section discusses experimental results. The aim is to reduce the feature set, resulting in accelerated training and enhanced model accuracy.

### A. Feature Importance

The feature importance analysis derived from RF and RFE identified the most relevant features for APT detection based on their contribution to the model's classification performance. Highly influential features are considered more relevant for APT detection [27].

The top 20 features selected by RFE were analyzed based on their contribution to the output classification. Features such as Flow Bytes/s and Bwd Packet Length Mean are consistent with known APT behavior, where exfiltration and beaconing cause abnormal traffic patterns.

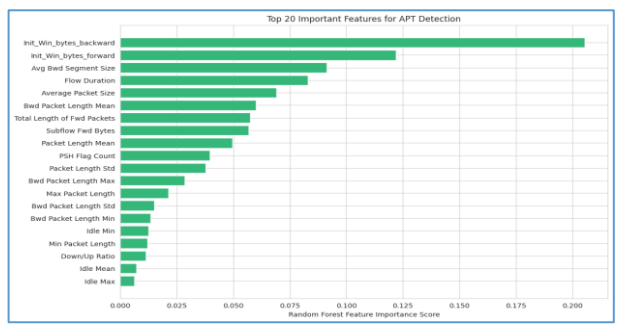

*Fig. 2. Feature Importance ranking*

The feature importance ranking shown in Fig. 2 illustrates the contribution of each feature to the decision-making process using RF model in the CICIDS2017 dataset. 20 features were selected from the original 80 features.

The most influential features were Init_Win_bytes_backward and Init_Win_bytes_forward, indicating their crucial role in distinguishing malicious traffic. Avg Bwd Segment Size and Flow Duration also showed high importance, suggesting that average backward segment size and the total duration of a network flow can reveal patterns associated with specific attack types. Features like Average Packet Size, Total Length of Fwd Packets, and Bwd Packet Length Mean provide additional context for packet structure and traffic flow. Lower contributions moderately to the model's predictions.

Features such as Idle Mean, Idle Max, and Down/Up Ratio have lower importance, indicating that while they add some value, their contribution is less impactful compared to the top-ranking features.

By leveraging RFE, the model has identified a subset of highly influential features, which help reduce dimensionality, enhance model performance, and improve interpretability. The insights from feature importance can guide cybersecurity analysts in focusing on critical traffic patterns for anomaly detection and refining IDS rules [32].

### B. Evaluation metrics

In ML paradigm, evaluation metrics measure APT detection performance. The metrics discussed are as follows:

- Accuracy is a metric finding the proportion of correct predictions made by a model. In simple terms, it gives a full picture of how well the model is performing.

$$Accuracy = \frac{TP + TN}{TP + TN + FP + FN} \quad (1)$$

- Precision measures the ratio of TP predictions among all positive predictions, which is crucial when FP can cause serious consequences [28].

$$Precision = \frac{TP}{TP + FP} \quad (2)$$

- Recall shows how well model can find all the actual positive instances. If recall is high, it means the model finds most or all the positive instances [29].

$$Recall = \frac{TP}{TP + FN} \quad (3)$$

- The F1-score uses the harmonic mean, which is valuable when balancing precision and recall [28].

$$F1 - Score = 2 \times \frac{Precision \times Recall}{Precision + Recall} \quad (4)$$

Table 2 presents RF model's performance results.

*Table 2. Our performance result of CICIDS2017 dataset*

| Parameter | RF-RFE |
|---|---|
| True Negatives (TN) | 503,567 |
| False Negatives (FN) | 110 |
| False Positives (FP) | 49 |
| True Positives (TP) | 747 |
| Accuracy | 0.9997 |

| Precision | 0.9384 |
|---|---|
| Recall | 0.8716 |
| F1-score | 0.9038 |

### C. Run time evaluation

Timely threat detection is important in Intrusion Detection Systems (IDS). The optimized feature set obtained through RFE led to a 40% reduction in training time and a 35% increase in inference efficiency, with an average prediction latency of approximately 20 ms per instance. This improvement underscores the model's suitability for deployment in real-time IDS environments, where low-latency processing and high responsiveness are crucial for effective cybersecurity defense.

### D. Explainable deep learning-based RF-RFE

In this study, we used SHAP to help us understand how the detection model makes decisions and makes the results easier to interpret. SHAP values quantify each feature's contribution to individual predictions, offering both global and local interpretability [30]. Feature importance rankings, summary plots, and decision plots are generated to visualize how specific attributes influence the model's output. This level of explainability is crucial for cybersecurity analysts, as it enables them to trace alerts back to specific traffic behaviors and enhances trust in the detection system [32].

We extract the top five feature scores from the RF model and present them in Fig. 3. Forward inter-arrival time standard deviation, active mean duration, and backward packets per second emerge as strong predictors. SHAP summary and dependence plots further illustrate how specific feature values, such as high variance in inter-arrival times or extended active durations, drive the model towards anomaly detection. These visual explanations offer both global and instance-level insights, empowering analysts to understand and trust each alert.

Fig. 3 presents a SHAP-based explanation for one test flow that our RF model classifies as Normal with 100% confidence (APT=0%). At the top, the prediction probabilities bar shows the model's final output.

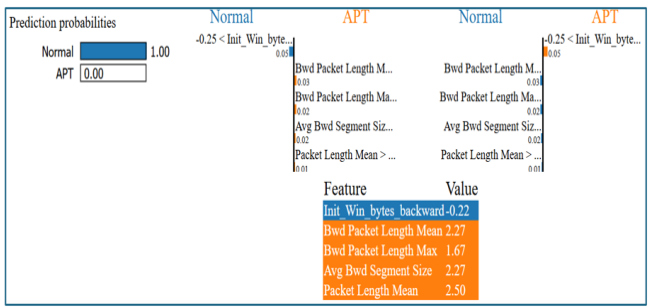

*Fig. 3. SHAP-based explanation*

For a normal classification, SHAP plot in Fig. 3 identified the five most influential features Init_Win_bytes_backward, Bwd Packet Length Mean, Bwd Packet Length Max, Avg Bwd Segment Size, and Packet Length Mean, alongside their numerical values. The four backward-traffic metrics strongly pull the model toward a normal verdict, while the slightly negative Init_Win_bytes_backward exerts only a minor push toward APT. By visualizing how each feature value influences the model's decision, SHAP provides transparent, instance-level interpretability, enabling analysts to understand precisely which flow characteristics determined the detection outcome [33].

Fig. 4 shows the SHAP-based global feature of importance, highlighting top features that help detect APT attacks. The top three (3) features, *Init_Win_bytes_backward, Flow Duration,* and *Init_Win_bytes_forward have* notably high mean |SHAP| values, indicating their strong predictive influence in distinguishing APT traffic patterns.

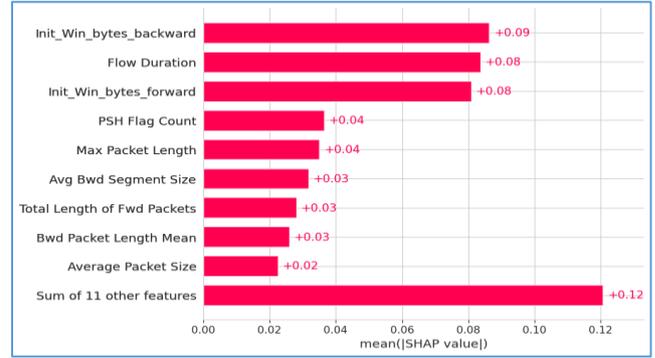

*Fig. 4. SHAP-based global feature importance for the RF classifier*

This suggests the model relies on these network flow characteristics to separate malicious APT traffic from benign behavior.

Fig. 5 presents a SHAP summary plot that illustrates both the magnitude and direction of each feature's impact on the RF model output. Features such as Init_Win_bytes_backward, Flow Duration, and Init_Win_bytes_forward consistently exert strong negative SHAP values, meaning higher values of these features generally push the model toward the "Normal" class, while lower values lean toward an "APT" prediction.

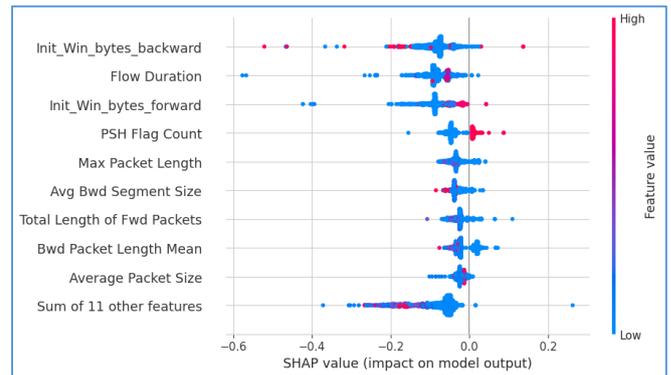

*Fig. 5. Top 20 Feature importance scores using Beeswarm plots on CICIDS2017*

Notably, the widespread SHAP values for these features reflect their strong discriminative power across different traffic flows. The visualization confirms their high average importance and how variations in specific feature values shift predictions on an instance-by-instance basis. By providing a detailed view of both feature importance and value gradients, the SHAP plot improves the interpretability of the model at a global level and supports clearer, more transparent decision-making in cybersecurity.

## IV. DISCUSSION

The study's findings underscore the effectiveness of the explainable RF-RFE framework in detecting APTs within complex network traffic. By systematically reducing the feature space, the framework enhances both training performance and model interpretability while maintaining high detection accuracy. The integration of SHAP further increases transparency, allowing analysts to trace model decisions back to specific network behaviors [33].

To assess model robustness, we evaluated the RF classifier on the CICIDS2017 dataset containing unseen traffic types such as Heartbleed and infiltration attacks. The model maintained high detection accuracy, showing resilience to novel attack vectors. However, like many supervised models, performance is dependent on the diversity of training data. Future iterations should incorporate adaptive learning or domain adaptation techniques to improve resilience against evolving threats.

*Table 3. Summary of related studies on feature selection*

| Authors | Dataset | Feature Selection | Performance |
|---|---|---|---|
| [6] | UNSWNB15 | IV-RFE | Reconnaissance: 91.6% accuracy |
| [9] | WUSTL-EHMS. | RFE with decision tree | Training accuracy: 99% Testing accuracy: 97.85% FAR: 0.03 |
| [10] | CICIDS2017, NSL-KDD | RFE, MI, and LFS | NSL-KDD all attacks: 99.95% accuracy |
| [11] | NSL-KDD | RFE, Correlation-based, Information Gain-based | Accuracy: 99.12%, F1-Score: 99.2% |
| [20] | NSL-KDD, UNSW-NB15, CSE–CIC–IDS2018 | Wrapper-FS with Decision Tree (J48) | Accuracy: 96.01%, DR: 96.20%, FAR: 9.57% |
| Our proposed framework | CICIDS2017 | Explainable RF-RFE | 20 top features on CICIDS2017: 99.9% accuracy |

## V. CONCLUSION

This study developed an interpretable Intrusion Detection System (IDS) capable of detecting Advanced Persistent Threats (APTs) with high accuracy. By integrating Recursive Feature Elimination (RFE) and Random Forest (RF), the framework efficiently reduced dimensionality and improved detection performance. SHapley Additive exPlanations (SHAP) was integrated to provide both global and instance-level interpretability, enabling transparent insight into the model's decision-making process. Experimental evaluation demonstrated that the system achieved a detection accuracy of 99.9% and exhibited robust performance.

Future work will evaluate the proposed RF-RFE framework in real-time deployment environments, where rapid response is crucial. Deep learning and ensemble-based models, such as Long Short-Term Memory (LSTM) networks can be explored to better capture temporal patterns in evolving cyber threats. These enhancements can improve the system's effectiveness and operational relevance in real-world intrusion detection scenarios. The framework will also be benchmarked against advanced classifiers, including LSTM, XGBoost, and generative AI-based techniques to compare performance in terms of accuracy, interpretability, and adaptability.

*Acknowledgement*
The authors would like to express their sincere gratitude to the ICCR-2025 reviewers and organizing committee for their valuable feedback and support throughout the review process. The research was conducted with the support of the Department of Artificial Intelligence, University of Malaya, as part of an ongoing effort to enhance explainable AI techniques in cybersecurity applications.

*References*

[1] Mutalib, N. H. A., Sabri, A. Q. M., Wahab, A. W. A., Abdullah, E. R. M. F., & AlDahoul, N. (2024). Explainable deep learning approach for advanced persistent threats (APTs) detection in cybersecurity: a review. Artificial Intelligence Review, 57(11), 297. https://doi.org/10.1007/s10462-024-10890-4

[2] Ferrag, M. A., Maglaras, L., Moschoyiannis, S., & Janicke, H. (2020). Deep learning for cyber security intrusion detection: Approaches, datasets, and comparative study. Journal of Information Security and Applications, 50, 102419. https://doi.org/https://doi.org/10.1016/j.jisa.2019.102419

[3] Xuan, C. D. (2021). Detecting APT Attacks Based on Network Traffic Using Machine Learning. JOURNAL OF WEB ENGINEERING, 20(1), 171–190. https://doi.org/10.13052/jwe1540-9589.2019

[4] Alsaffar, A. M., Nouri-Baygi, M., & Zolbanin, H. (2024). Enhancing Intrusion Detection Systems with Dimensionality Reduction and Multi-Stacking Ensemble Techniques. Algorithms, 17(12). https://doi.org/10.3390/a17120550

[5] Yaqoob, A., Verma, N.K., Mir, M.A. et al. SGA-Driven feature selection and random forest classification for enhanced breast cancer diagnosis: A comparative study. Sci Rep 15, 10944 (2025). https://doi.org/10.1038/s41598-025-95786-1

[6] Barbosa, G. N. N., Andreoni, M., & Mattos, D. M. F. (2024). Optimizing feature selection in intrusion detection systems: Pareto dominance set approaches with mutual information and linear correlation. Ad Hoc Networks, 159, 103485. https://doi.org/https://doi.org/10.1016/j.adhoc.2024.103485

[7] T, S., & E A, M. A. (2025). A novel stable feature selection algorithm for machine learning based intrusion detection system. *Procedia Computer Science*, 252, 738–747. https://doi.org/https://doi.org/10.1016/j.procs.2025.01.034

[8] Lan Wang, Jialing Xu, Liyun Jia, Tao Wang, Yujie Xu, Xingchen Liu, Multi-strategy RIME optimization algorithm for feature selection of network intrusion detection,Computers & Security,Volume 153,2025, 104393, ISSN 0167-4048, https://doi.org/10.1016/j.cose.2025.104393.

[9] Qasem, A. A., Qutqut, M. H., Alhaj, F., & Kitana, A. (2024). SRFE: A stepwise recursive feature elimination approach for network intrusion detection systems. Peer-to-Peer Networking and Applications, 17(6), 3634–3649. https://doi.org/10.1007/s12083-024-01763-2

[10] Albasheer Mohamed, F. O., & Agarwal, M. (2024). Using Recursive Feature Elimination Feature Selection based Machine Learning Classifier for Attack Classification on UNSW-NB 15 dataset. 2024 IEEE 9th International Conference for Convergence in Technology (I2CT), 1–7. https://doi.org/10.1109/I2CT61223.2024.10544076

[11] Lazrek, G., Chetioui, K., Balboul, Y., Mazer, S., & El bekkali, M. (2024a). An RFE/Ridge-ML/DL based anomaly intrusion detection approach for securing IoMT system. Results in Engineering, 23, 102659. https://doi.org/https://doi.org/10.1016/j.rineng.2024.102659

[12] Urmi, W. F., Uddin, M. N., Uddin, M. A., Talukder, Md. A., Hasan, Md. R., Paul, S., Chanda, M., Ayoade, J., Khraisat, A., Hossen, R., & Imran, F. (2024). A stacked ensemble approach to detect cyber attacks based on feature selection techniques. International Journal of

Cognitive Computing in Engineering, 5, 316–331. https://doi.org/https://doi.org/10.1016/j.ijcce.2024.07.005

[13] Sharma, A., & Singh, M. (2024). Batch reinforcement learning approach using recursive feature elimination for network intrusion detection. Engineering Applications of Artificial Intelligence, 136, 109013. https://doi.org/https://doi.org/10.1016/j.engappai.2024.109013

[14] Khan, A. A., Chaudhari, O., & Chandra, R. (2024). A review of ensemble learning and data augmentation models for class imbalanced problems: Combination, implementation and evaluation. Expert Systems with Applications, 244, 122778. https://doi.org/https://doi.org/10.1016/j.eswa.2023.122778

[15] Xia, S., & Yang, Y. (2023). A Model-Free Feature Selection Technique of Feature Screening and Random Forest-Based Recursive Feature Elimination. International Journal of Intelligent Systems, 2023, 1–16. https://doi.org/10.1155/2023/2400194

[16] Lian, W., Nie, G., Jia, B., Shi, D., Fan, Q., & Liang, Y. (2020). An Intrusion Detection Method Based on Decision Tree-Recursive Feature Elimination in Ensemble Learning. Mathematical Problems in Engineering, 2020, 1–15. https://doi.org/10.1155/2020/2835023

[17] Pes, B. (2021). Learning from High-Dimensional and Class-Imbalanced Datasets Using Random Forests. Information, 12(8). https://doi.org/10.3390/info12080286

[18] Barbieri, M. C., Grisci, B. I., & Dorn, M. (2024). Analysis and comparison of feature selection methods towards performance and stability. Expert Systems with Applications, 249, 123667. https://doi.org/https://doi.org/10.1016/j.eswa.2024.123667

[19] Patil, S., Varadarajan, V., Mazhar, S. M., Sahibzada, A., Ahmed, N., Sinha, O., Kumar, S., Shaw, K., & Kotecha, K. (2022). Explainable Artificial Intelligence for Intrusion Detection System. Electronics, 11(19), 3079. https://doi.org/10.3390/electronics11193079

[20] Logeswari, G., Thangaramya, K., Selvi, M., & Roselind, J. D. (2025). An improved synergistic dual-layer feature selection algorithm with two type classifier for efficient intrusion detection in IoT environment. Scientific Reports, 15(1), 8050. https://doi.org/10.1038/s41598-025-91663-z

[21] Umar, M. A., Chen, Z., Shuaib, K., & Liu, Y. (2025). Effects of feature selection and normalization on network intrusion detection. Data Science and Management, 8(1), 23–39. https://doi.org/https://doi.org/10.1016/j.dsm.2024.08.001

[22] A. Adadi and M. Berrada, "Peeking Inside the Black-Box: A Survey on Explainable Artificial Intelligence (XAI)," in IEEE Access, vol. 6, pp. 52138-52160, 2018, doi: 10.1109/ACCESS.2018.2870052.

[23] Arreche, O., Guntur, T., & Abdallah, M. (2024). XAI-IDS: Toward Proposing an Explainable Artificial Intelligence Framework for Enhancing Network Intrusion Detection Systems. Applied Sciences, 14(10), 4170. https://doi.org/10.3390/app14104170

[24] Udofot, A. I., Oluseyi, O. M., & Bassey, E. Explainable AI for cyber security. Improving transparency and trust in intrusion detection systems.

[25] Panigrahi, R., & Borah, S. (2018). A detailed analysis of CICIDS2017 dataset for designing Intrusion Detection Systems. International Journal of Engineering & Technology, 7, 479–482.

[26] Z. A. E. Houda, B. Brik and L. Khoukhi, ""Why Should I Trust Your IDS?": An Explainable Deep Learning Framework for Intrusion Detection Systems in Internet of Things Networks," in IEEE Open Journal of the Communications Society, vol. 3, pp. 1164-1176, 2022, doi: 10.1109/OJCOMS.2022.3188750.

[27] Jaw, E., & Wang, X. (2021). Feature Selection and Ensemble-Based Intrusion Detection System: An Efficient and Comprehensive Approach. Symmetry, 13(10), 1764. https://doi.org/10.3390/sym13101764

[28] Shoukat, S., Gao, T., Javeed, D., Saeed, M. S., & Adil, M. (2025). Trust my IDS: An explainable AI integrated deep learning-based transparent threat detection system for industrial networks. Computers & Security, 149, 104191. https://doi.org/https://doi.org/10.1016/j.cose.2024.10419

[29] AlDahoul, N., Abdul Karim, H. & Ba Wazir, A.S. Model fusion of deep neural networks for anomaly detection. J Big Data 8, 106 (2021). https://doi.org/10.1186/s40537-021-00496-w

[30] T. Nakanishi, P. Chophuk and K. Chinnasarn, "Evolving Feature Selection: Synergistic Backward and Forward Deletion Method Utilizing Global Feature Importance," in IEEE Access, vol. 12, pp. 88696-88714, 2024, doi: 10.1109/ACCESS.2024.3418499.

[31] Mao, J., Yang, X., Hu, B., Lu, Y., & Yin, G. (2025). Intrusion Detection System Based on Multi-Level Feature Extraction and Inductive Network. Electronics, 14(1). https://doi.org/10.3390/electronics14010189

[32] Wang, H., Liang, Q., Hancock, J.T. et al. Feature selection strategies: a comparative analysis of SHAP-value and importance-based methods. J Big Data 11, 44 (2024). https://doi.org/10.1186/s40537-024-00905-w

[33] Li, J., Meng, X., Qi, Z. et al. Attack stage detection method based on vector reconstruction error autoencoder and explainable artificial intelligence. J Supercomput 81, 62 (2025). https://doi.org/10.1007/s11227-024-06473-3